\documentclass{aa}
\usepackage{graphics}
\usepackage{subfigure}
\usepackage{latexsym,amssymb}

\begin{document}

\thesaurus{11.01.2;              
           11.02.2 PKS 2155-304; 
           11.16.1;              
           13.09.2}              

\title{ISO observations of the BL Lac object 
PKS 2155--304\thanks{Based on observations with ISO, an ESA project 
with instruments funded by ESA Member States (especially the PI countries: 
France, Germany, the Netherlands and the United Kingdom) with the 
participation of ISAS and NASA.}.}

\author{E. Bertone\inst{1,2}, G. Tagliaferri\inst{1}, G. Ghisellini\inst{1},
A. Treves\inst{3}, P. Barr\inst{4}, A. Celotti\inst{5}, M. Chiaberge\inst{5}, 
L.~Maraschi\inst{6}}

\institute{Osservatorio Astronomico di Brera, via Bianchi 46, I--23807 Merate
(LC), Italy
\and Dipartimento di Fisica, Universit\`a degli Studi di Milano, via Celoria
16, I--20133 Milano, Italy
\and Universit\`a dell'Insubria, via Lucini 3, I--22100 Como, Italy
\and Astrophysics Division, ESA, ESTEC, Postbus 299, 2200 AG Noordwijk, The Netherlands
\and SISSA, via Beirut 2-4, I-34014 Trieste, Italy
\and Osservatorio Astronomico di Brera, via Brera 28, I--20121 Milano, Italy}

\offprints{bertone@merate.mi.astro.it}

\titlerunning{ISO observations of PKS 2155--304}

\authorrunning{Bertone et al.}

\maketitle


\begin{abstract}

The BL Lacertae object PKS 2155--304 was observed by the Infrared Space
Observatory in May and June 1996, during a multiwavelength campaign.
These are the first observations in the mid-- and far--infrared bands 
since IRAS.
In the observing period, the source showed no detectable time variability at
4.0, 14.3, 60 and 90 $\mu$m.
The spectrum from 2.8 to 100 $\mu$m is well fitted by a single power law with 
energy spectral index $\alpha = 0.4$, intermediate between the flatter
radio spectrum and the steeper simultaneous optical spectrum.
The overall infrared to X--ray spectral energy distribution 
can be well explained by optically thin synchrotron emission, with
negligible contributions from thermal sources. 
We also show that the host galaxy flux is negligible in this
spectral range.


\end{abstract}

\section{Introduction}

\bigskip
BL Lacertae objects are characterized by an intense and variable non--thermal 
continuum, that extends from the radio to the gamma--ray band. 
This is commonly attributed to synchrotron and inverse Compton radiation
from a relativistic jet pointing toward the observer (see 
\cite{UlrichMaraschiUrry} for a review).
In a $\nu F_{\nu}$ representation, their overall spectrum has two broad peaks,
one at low energies (IR--X) due to synchrotron radiation and one at
higher energies (X--$\gamma$), plausibly due to inverse Compton scattering.

PKS 2155--304 is one of the brightest BL Lacs from the optical to 
the X--ray band
with the synchrotron peak in the UV- soft X--ray range, corresponding
to the definition of High frequency peak BL Lac objects 
(HBL) (\cite{PadovaniGiommi}), which have the synchrotron peak 
at the highest frequencies, low luminosity and a small ratio between the 
$\gamma$--ray and the synchrotron peak luminosities.
The gamma-ray spectrum is flat ($\alpha_{\gamma} \simeq 0.7$\footnote{$\alpha$
is defined as $F(\nu) \propto \nu^{-\alpha}$.} in the 0.1-10 GeV
energy range), indicating that the Compton peak is beyond $\sim 10$ GeV.
Recently it has been detected in the TeV band (Chadwick et
al. \cite{Chadwick}).
Due to these characteristics, PKS\,2155--304 has been the target of numerous 
multiwavelength campaigns (e.g. Edelson et al.\cite{Edelson} for November
1991, Urry et al. \cite{Urry97} for May 1994).
The study of the simultaneous behavior of the source at different 
frequencies is important 
in order to understand the emission mechanisms and to constrain the physical 
properties of the emitting region.

In 1996 May--June, an intense multiwavelength monitoring was carried out 
involving optical telescopes, UV, X--ray and $\gamma$--ray satellites.
Thanks to the Infrared Space Observatory (ISO), for
the first time we had infrared simultaneous observations. These are the
first observations of this object in the mid-- and far-- infrared since 
IRAS. PKS 2155--304 was detected by IRAS in 1983 at 12, 25, 60 microns
with a flux of about 100 mJy in all three bands (Impey \& Neugebauer
\cite{ImpeyNeugebauer}).
In this object the IR emission is at frequencies lower than the synchrotron
peak, and the spectral shape in this band can reveal if there are relevant
thermal contributions (e.g. by the host galaxy or by a dusty torus around the
nucleus) or if the emission can be entirely attributed to synchrotron 
radiation.

Here we present the ISO observations of PKS 2155--304, carried out during the
campaign in 1996 May--June, covering a wavelength range from 2.8 to 200 
$\mu$m.
This is complemented by some simultaneous BVR observations from the 
Dutch 0.9 m ESO
telescope. Results from ISO observations of 1996 November and 1997 May are
also reported.

The paper is organized as follows:
a brief description of the ISO instruments and of the observations are
given in section \ref{sec:ISOobs} and the results are reported in section
\ref{sec:ISOresults}. In section \ref{sec:optobs} we present the optical data
and in section \ref{sec:discussion} we compare our results with the
theoretical models. PKS 2155--304 is a weak IR source for ISO. Therefore
considerable care was taken in data reduction and background
subtraction. Details are given in Appendix A.


\section{ISO observations}
\label{sec:ISOobs}

PKS 2155--304 was observed with ISO between 
1996 May 7 and June 8. Two additional observations were performed 
on 1996 November 23 and 1997 May 15.

The ISO satellite (\cite{Kessler}) is equipped with a 60 cm
Ritchey--Chr\'etien telescope and has four scientific instruments on board.
For the PKS 2155--304 observations both the camera ISOCAM and the
photometer ISOPHOT were used.

The 32x32 pixel imaging camera ISOCAM (\cite{Cesarsky}) has two detectors:
an InSb CID (Charge Injection Device) 
for short wavelengths (SW detector; 2.5 -- 5.5 $\mu$m) and a Si:Ga
photoconductor array for longer wavelengths (LW detector; 4 -- 17 $\mu$m). 
It is equipped with a set of 21 broad--band filters and a circular variable 
filter with a higher spectral resolution. The spatial resolution ranges from 
1.5\arcsec\,to 12\arcsec\,per pixel.

The photometer ISOPHOT (\cite{Lemke}) has three subsystems: a photo--polarimeter
(PHT--P) (3 -- 120 $\mu$m), which has 3 detectors, sensitive at different 
wavelengths, 14 broad-band filters and different apertures, from 5\arcsec\,to 
180\arcsec; an imaging photometric camera (PHT--C) (50 -- 240 $\mu$m),
with a 3x3 and a 2x2 pixel detectors, a field of view of
43.5\arcsec x43.5\arcsec\,and 89.4\arcsec  x89.4\arcsec\,per pixel, 
respectively, and 11 broad--band filters; two low--resolution grating 
spectrometers (PHT--S) (2.5 -- 5 $\mu$m and 6 -- 12 $\mu$m).

In order to determine the variability characteristics in the infrared band,
15 identical observations were performed in the period between 1996 May 7 
and June 8, at 4.0, 14.3, 60, 90 and 170 $\mu$m (see Tab. \ref{tab:filters} 
for the filter characteristics). From May 13 to May 27 ISO observed 
PKS 2155--304 almost each day.
The observing modes (AOTs, Astronomical Observation Templates) 
were CAM01 (\cite{ISOCAMobsman}), in single pointing mode, and PHT22
(\cite{ISOPHOTobsman}), in rectangular chopped mode (see Appendix
\ref{sec:PHOTdatareduction}).

\begin{table}[t]
\caption{Characteristics of the filters used with ISOCAM and ISOPHOT.}
\begin{tabular}{lr@{}lr@{--}lc}
\noalign{\smallskip}
\hline
\noalign{\smallskip}
filter & \multicolumn{2}{c}{ref. $\lambda$ ($\mu$m)} &
\multicolumn{2}{c}{range ($\mu$m)} & $\lambda/ \Delta\lambda$ \\
\noalign{\smallskip}
\hline
\noalign{\smallskip}
SW4     &   2&.8  & 2.50&3.05 &  5   \\
SW2     &   3&.3  & 3.20&3.40 & 17   \\
SW6     &   3&.7  & 3.45&4.00 &  7   \\
SW5     &   4&.0  & 3.00&5.50 &  2   \\
SW11    &   4&.26 & 4.16&4.37 & 20   \\
SW10    &   4&.6  & 4.53&4.88 & 13   \\
LW4     &   6&.0  & 5.50&6.50 &  6   \\
LW6     &   7&.7  & 7.00&8.50 &  5   \\
LW7     &   9&.6  & 8.50&10.7 &  4   \\
LW8     &  11&.3  & 10.7&12.0 &  9   \\
LW3     &  14&.3  & 12.0&18.0 &  3   \\
LW9     &  14&.9  & 14.0&16.0 &  9   \\
P2\_25  &  25&    & 19.2&28.4 &  2.5   \\
C1\_60  &  60&    & 49  &63   &  2.5   \\
C1\_70  &  80&    & 55  &105  &  2.5   \\
C1\_90  &  90&    & 69  &121  &  1.9   \\
C1\_100 & 100&    & 82  &125  &  2.4   \\
C2\_160 & 170&    & 129 &219  &  2     \\
C2\_180 & 180&    & 150 &211  &  2.6   \\
C2\_200 & 200&    & 171 &238  &  3     \\
\noalign{\smallskip}
\hline
\end{tabular}
\label{tab:filters}
\end{table}

On 1996 May 27 the source was observed in a large wavelength range (from 2.8
to 200 $\mu$m) with 17 different filters in order to determine the infrared 
spectrum.
The same AOTs as before were used, except the 
observation with the P2\_25 filter, for which the PHT03, still in 
rectangular chopped mode, was used.

On 1997 May 15 two 3x3 raster scans, centered on PKS 2155--304
(R.A. 21h 58m 52s, Dec --30\degr \, 13\arcmin \, 32\arcsec) were performed with
the photometric camera PHT--C, at 60 $\mu$m and at 180 $\mu$m; the distance 
between two adjacent raster positions was
180\arcsec, in order to have an almost complete sky coverage of an area
of 9\arcmin\,side. This mapping was performed to search for any
structure in the cirrus clouds; a non flat background could compromise a
reliable photometry of the source. 
In this observation the AOT PHT22 was used in staring mode.

The ISOPHOT observation of 1996 May 25 failed because of problems
during the instrument activation.

The complete log of the observations is shown in Tabs. \ref{tab:CAMobslog} and 
\ref{tab:PHOTobslog}.

\begin{table*}[t]
\caption{ISOCAM observation log.}
\begin{tabular}{lllrrr|lllrrr}
\noalign{\smallskip}
\hline
\noalign{\smallskip}
\multicolumn{2}{c}{obs. time} & filter & \multicolumn{1}{c}{$\lambda$} &
pfov &  frames & 
\multicolumn{2}{c}{obs. time} & filter & \multicolumn{1}{c}{$\lambda$} &
pfov &  \multicolumn{1}{l}{frames}\\
yy/mm/dd & mjd--50000 & &  $\mu$m   &  \multicolumn{1}{c}{\arcsec} & \# &
yy/mm/dd & mjd--50000 & &  $\mu$m   &  \multicolumn{1}{c}{\arcsec} & \multicolumn{1}{r}{\#}\\
\noalign{\smallskip}
\hline
\noalign{\smallskip}
96/05/07 & 210.9703 & SW5 &  4.0 & 6.0 &   59 &
96/05/25 & 228.9254 & SW5 &  4.0 & 6.0 &   59 \\
96/05/07 & 210.9715 & LW3 & 14.3 & 6.0 &   41 &
96/05/25 & 228.9266 & LW3 & 14.3 & 6.0 &   41 \\
96/05/13 & 216.9549 & SW5 &  4.0 & 6.0 &   59 &
96/05/26 & 229.9227 & SW5 &  4.0 & 6.0 &   59 \\
96/05/13 & 216.9564 & LW3 & 14.3 & 6.0 &   41 &
96/05/26 & 229.9240 & LW3 & 14.3 & 6.0 &   41 \\
96/05/15 & 218.9896 & SW5 &  4.0 & 6.0 &   59 &
96/05/27 & 230.9207 & SW4 &  2.8 & 3.0 &  159 \\
96/05/15 & 218.9909 & LW3 & 14.3 & 6.0 &   41 &
96/05/27 & 230.9243 & SW2 &  3.3 & 6.0 &  110 \\
96/05/16 & 219.9621 & SW5 &  4.0 & 6.0 &   59 &
96/05/27 & 230.9270 & SW6 &  3.7 & 6.0 &  111 \\
96/05/16 & 219.9634 & LW3 & 14.3 & 6.0 &   40 &
96/05/27 & 230.9299 & SW11 &  4.26 & 6.0 &  162 \\
96/05/18 & 221.0461 & SW5 &  4.0 & 6.0 &   59 &
96/05/27 & 230.9337 & SW10 &  4.6 & 6.0 &  162 \\
96/05/18 & 221.0473 & LW3 & 14.3 & 6.0 &   42 &
96/05/27 & 230.9376 & LW4 &  6.0 & 3.0 &   59 \\
96/05/18 & 221.9426 & SW5 &  4.0 & 6.0 &   59 &
96/05/27 & 230.9391 & LW6 &  7.7 & 3.0 &   87 \\
96/05/18 & 221.9438 & LW3 & 14.3 & 6.0 &   41 &
96/05/27 & 230.9413 & LW7 &  9.6 & 3.0 &   87 \\
96/05/19 & 222.9401 & SW5 &  4.0 & 6.0 &   59 &
96/05/27 & 230.9435 & LW8 & 11.3 & 3.0 &   87 \\
96/05/19 & 222.9414 & LW3 & 14.3 & 6.0 &   41 &
96/05/27 & 230.9462 & LW9 & 14.9 & 3.0 &  114 \\
96/05/21 & 224.0848 & SW5 &  4.0 & 6.0 &   59 &
96/06/04 & 238.0014 & SW5 &  4.0 & 6.0 &   59 \\
96/05/21 & 224.0897 & LW3 & 14.3 & 6.0 &   41 &
96/06/04 & 238.0026 & LW3 & 14.3 & 6.0 &   41 \\
96/05/21 & 224.9352 & SW5 &  4.0 & 6.0 &   59 &
96/06/08 & 242.8894 & SW5 &  4.0 & 6.0 &   59 \\
96/05/21 & 224.9365 & LW3 & 14.3 & 6.0 &   41 &
96/06/08 & 242.8906 & LW3 & 14.3 & 6.0 &   40 \\
96/05/23 & 226.0251 & SW5 &  4.0 & 6.0 &   60 &
96/11/23 & 410.4845 & SW5 &  4.0 & 6.0 &   59 \\
96/05/23 & 226.0263 & LW3 & 14.3 & 6.0 &   40 &
96/11/23 & 410.4857 & LW3 & 14.3 & 6.0 &   40 \\
96/05/24 & 227.9279 & SW5 &  4.0 & 6.0 &   59 &
97/05/15 & 583.4535 & SW5 &  4.0 & 6.0 &   59 \\
96/05/24 & 227.9291 & LW3 & 14.3 & 6.0 &   41 &
97/05/15 & 583.4548 & LW3 & 14.3 & 6.0 &   41 \\
\noalign{\smallskip}
\hline
\noalign{\smallskip}
\multicolumn{12}{l}{\footnotesize Note. The integration time of each frame is 2.1 s; the gain is 2.} \\
\end{tabular}
\label{tab:CAMobslog}
\end{table*}

\begin{table*}
\caption{ISOPHOT observation log.}
\begin{tabular}{lllrllr|lllrllr}
\noalign{\smallskip}
\hline
\noalign{\smallskip}
\multicolumn{2}{c}{obs. time} & filter & \multicolumn{1}{c}{$\lambda$} & 
pfov & $t_{int}$ &  $t_{tot}$ &
\multicolumn{2}{c}{obs. time} & filter & \multicolumn{1}{c}{$\lambda$} & 
pfov & $t_{int}$ &  $t_{tot}$ \\
yy/mm/dd & mjd--50000 & &  $\mu$m   &  \multicolumn{1}{c}{\arcsec} & s & s &
yy/mm/dd & mjd--50000 & &  $\mu$m   &  \multicolumn{1}{c}{\arcsec} & s & s \\
\noalign{\smallskip}
\hline
\noalign{\smallskip}
96/05/07 & 210.9736 & C2\_160 & 170 & 89.4 & 4 & 128 & 
96/05/23 & 226.0347 & C1\_90  & 90  & 43.5 & 1 & 64   \\
96/05/07 & 210.9778 & C1\_60  & 60  & 43.5 & 2 & 256 & 
96/05/24 & 227.9313 & C2\_160 & 170 & 89.4 & 4 & 128  \\
96/05/07 & 210.9798 & C1\_90  & 90  & 43.5 & 1 & 64  & 
96/05/24 & 227.9355 & C1\_60  & 60  & 43.5 & 2 & 256  \\
96/05/13 & 216.9585 & C2\_160 & 170 & 89.4 & 4 & 128 & 
96/05/24 & 227.9375 & C1\_90  & 90  & 43.5 & 1 & 64   \\
96/05/13 & 216.9627 & C1\_60  & 60  & 43.5 & 2 & 256 & 
96/05/26 & 229.9261 & C2\_160 & 170 & 89.4 & 4 & 128  \\
96/05/13 & 216.9648 & C1\_90  & 90  & 43.5 & 1 & 64  & 
96/05/26 & 229.9303 & C1\_60  & 60  & 43.5 & 2 & 256  \\
96/05/15 & 218.9930 & C2\_160 & 170 & 89.4 & 4 & 128 &
96/05/26 & 229.9323 & C1\_90  & 90  & 43.5 & 1 & 64   \\
96/05/15 & 218.9972 & C1\_60  & 60  & 43.5 & 2 & 256 &
96/05/27 & 230.9556 & P2\_25  & 25  & 52.0 & 8 & 1024  \\
96/05/15 & 218.9992 & C1\_90  & 90  & 43.5 & 1 & 64  &
96/05/27 & 230.9691 & C2\_200 & 200 & 89.4 & 16 & 1024  \\
96/05/16 & 219.9655 & C2\_160 & 170 & 89.4 & 4 & 128 &
96/05/27 & 230.9767 & C2\_160 & 170 & 89.4 & 4 & 256  \\
96/05/16 & 219.9697 & C1\_60  & 60  & 43.5 & 2 & 256 &
96/05/27 & 230.9809 & C1\_100 & 100 & 43.5 & 2 & 128  \\
96/05/16 & 219.9718 & C1\_90  & 90  & 43.5 & 1 & 64  &
96/05/27 & 230.9832 & C1\_70  & 80  & 43.5 & 2 & 256  \\
96/05/18 & 221.0495 & C2\_160 & 170 & 89.4 & 4 & 128 &
96/05/27 & 230.9878 & C1\_60  & 60  & 43.5 & 2 & 512  \\
96/05/18 & 221.0536 & C1\_60  & 60  & 43.5 & 2 & 256 &
96/05/27 & 230.9918 & C1\_90  & 90  & 43.5 & 1 & 128  \\
96/05/18 & 221.0557 & C1\_90  & 90  & 43.5 & 1 & 64  &
96/06/04 & 238.0047 & C2\_160 & 170 & 89.4 & 4 & 128  \\
96/05/18 & 221.9460 & C2\_160 & 170 & 89.4 & 4 & 128 &
96/06/04 & 238.0089 & C1\_60  & 60  & 43.5 & 2 & 256  \\
96/05/18 & 221.9502 & C1\_60  & 60  & 43.5 & 2 & 256 &
96/06/04 & 238.0109 & C1\_90  & 90  & 43.5 & 1 & 64   \\
96/05/18 & 221.9522 & C1\_90  & 90  & 43.5 & 1 & 64  &
96/06/08 & 242.8927 & C2\_160 & 170 & 89.4 & 4 & 128  \\
96/05/19 & 222.9435 & C2\_160 & 170 & 89.4 & 4 & 128 &
96/06/08 & 242.8969 & C1\_60  & 60  & 43.5 & 2 & 256  \\
96/05/19 & 222.9477 & C1\_60  & 60  & 43.5 & 2 & 256 &
96/06/08 & 242.9024 & C1\_90  & 90  & 43.5 & 1 & 64   \\
96/05/19 & 222.9498 & C1\_90  & 90  & 43.5 & 1 & 64  & 
96/11/23 & 410.4878 & C2\_160 & 170 & 89.4 & 4 & 128  \\
96/05/21 & 224.0918 & C2\_160 & 170 & 89.4 & 4 & 128 &
96/11/23 & 410.4920 & C1\_60  & 60  & 43.5 & 2 & 256  \\
96/05/21 & 224.0960 & C1\_60  & 60  & 43.5 & 2 & 256 &
96/11/23 & 410.4940 & C1\_90  & 90  & 43.5 & 1 & 64   \\
96/05/21 & 224.0981 & C1\_90  & 90  & 43.5 & 1 & 64  &
97/05/15 & 583.4569 & C2\_160 & 170 & 89.4 & 4 & 128  \\
96/05/21 & 224.9386 & C2\_160 & 170 & 89.4 & 4 & 128 &
97/05/15 & 583.4611 & C1\_60  & 60  & 43.5 & 2 & 256  \\
96/05/21 & 224.9428 & C1\_60  & 60  & 43.5 & 2 & 256 &
97/05/15 & 583.4631 & C1\_90  & 90  & 43.5 & 1 & 64   \\
96/05/21 & 224.9448 & C1\_90  & 90  & 43.5 & 1 & 64  &
97/05/15 & 583.4741 & C1\_60  & 60  & 43.5 & 2 & 1260 \\
96/05/23 & 226.0291 & C2\_160 & 170 & 89.4 & 4 & 128 &
97/05/15 & 583.4920 & C1\_180 & 180 & 89.4 & 4 & 1260 \\
96/05/23 & 226.0326 & C1\_60  & 60  & 43.5 & 2 & 256 &
         &          &         &     &      &   &      \\
\noalign{\smallskip}
\hline
\noalign{\smallskip}
\multicolumn{14}{l}{\footnotesize Note. The two last observations are 3 x 3 raster scans.} \\
\end{tabular}
\label{tab:PHOTobslog}
\end{table*}


\section{ISO results}
\label{sec:ISOresults}

\subsection{The light curves}

\begin{figure}[t]
\resizebox{\hsize}{!}{\includegraphics{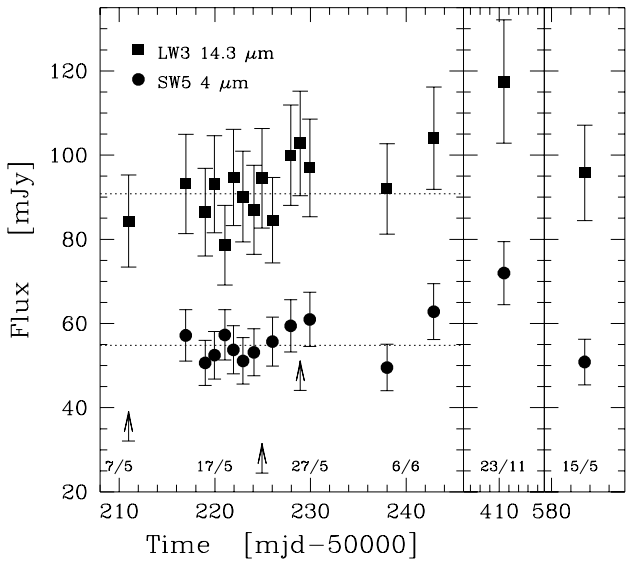}}
\caption{ISOCAM light curves of PKS 2155--304. The dotted curves represent the 
fitted constant of the best sampled period, from 1996 May 13 to May
27. The upward arrows are the lower limits in the SW5 curve.}
\label{fig:lcisocam}
\end{figure}

The data and the corresponding light curves at 4.0 
(SW5 filter), 14.3 (LW3), 60 (C1\_60)
and 90 $\mu$m (C1\_90) are reported in
Tabs. \ref{tab:lcisocam} and \ref{tab:lcisophot} and shown in Figs. 
\ref{fig:lcisocam} and \ref{fig:lcisophot}. 
The discussion on the data analysis and error evaluation is 
given in Appendix A.
At 170 $\mu$m (C2\_160), the 
source is not detected: the three sigma upper limit at this wavelength is 
1235 mJy (see Fig. \ref{fig:spectrum}). 

\begin{table}[th]
\caption{ISOCAM light curves.}
\begin{tabular}{llr@{}lr}
\multicolumn{5}{c}{SW5 filter (4.0 $\mu$m)}  \\
\noalign{\smallskip}
\hline
\noalign{\smallskip}
\multicolumn{2}{c}{obs. time} &
\multicolumn{2}{c}{flux} & 
\multicolumn{1}{c}{flux$_{AAR}^{(a)}$} \\
yy/mm/dd & mjd--50000 & \multicolumn{2}{c}{mJy} & 
\multicolumn{1}{c}{mJy} \\ 
\noalign{\smallskip}
\hline
\noalign{\smallskip}
96/05/05  & 210.9703 &  32&.1  $^{(b)}$ &  35.4 $\pm$  1.5 \\
96/05/13  & 216.9549 &  57&.2 $\pm$ 6.1 &  51.8 $\pm$  1.9 \\
96/05/15  & 218.9896 &  50&.6 $\pm$ 5.4 &  54.1 $\pm$  1.9 \\
96/05/16  & 219.9621 &  52&.5 $\pm$ 5.7 &  49.8 $\pm$  2.0 \\
96/05/18  & 221.0461 &  57&.3 $\pm$ 6.0 &  56.7 $\pm$  1.7 \\
96/05/18  & 221.9426 &  53&.8 $\pm$ 5.7 &  48.9 $\pm$  1.8 \\
96/05/19  & 222.9401 &  51&.1 $\pm$ 5.5 &  47.0 $\pm$  1.9 \\
96/05/21  & 224.0848 &  53&.2 $\pm$ 5.6 &  55.3 $\pm$  1.8 \\
96/05/21  & 224.9352 &  24&.5  $^{(b)}$ &  21.4 $\pm$  1.5 \\
96/05/23  & 226.0251 &  55&.7 $\pm$ 5.8 &  58.5 $\pm$  1.7 \\
96/05/24  & 227.9279 &  59&.4 $\pm$ 6.2 &  60.2 $\pm$  1.9 \\
96/05/25  & 228.9254 &  44&.1  $^{(b)}$ &  47.8 $\pm$  2.0 \\
96/05/26  & 229.9227 &  61&.0 $\pm$ 6.4 &  58.4 $\pm$  2.0 \\
96/06/04  & 238.0014 &  49&.5 $\pm$ 5.5 &  39.4 $\pm$  1.9 \\
96/06/08  & 242.8894 &  62&.8 $\pm$ 6.6 &  60.1 $\pm$  2.0 \\
96/11/23  & 410.4845 &  72&.0 $\pm$ 7.5 &  75.4 $\pm$  2.1 \\
97/05/15  & 583.4535 &  50&.9 $\pm$ 5.4 &  43.4 $\pm$  1.6 \\
\noalign{\smallskip}
\hline
\noalign{\smallskip}
\\
\multicolumn{5}{c}{LW3 filter (14.3 $\mu$m)}  \\ 
\noalign{\smallskip}
\hline
\noalign{\smallskip}
\multicolumn{2}{c}{obs. time} &
\multicolumn{2}{c}{flux} &
\multicolumn{1}{c}{flux$_{AAR}^{(a)}$} \\
yy/mm/dd & mjd--50000 & \multicolumn{2}{c}{mJy} & 
\multicolumn{1}{c}{mJy}    \\ 
\noalign{\smallskip}
\hline
\noalign{\smallskip}
96/05/05  & 210.9715 &  84&.3 $\pm$ 10.9 &  66.8 $\pm$  5.5 \\
96/05/13  & 216.9564 &  93&.2 $\pm$ 11.8 & 100.3 $\pm$  7.8 \\
96/05/15  & 218.9909 &  86&.5 $\pm$ 10.4 &  77.5 $\pm$  5.2 \\
96/05/16  & 219.9634 &  93&.1 $\pm$ 11.5 &  80.6 $\pm$  5.9 \\
96/05/18  & 221.0473 &  78&.6 $\pm$  9.4 &  90.0 $\pm$  6.0 \\
96/05/18  & 221.9438 &  94&.7 $\pm$ 11.4 &  91.8 $\pm$  6.2 \\
96/05/19  & 222.9414 &  90&.1 $\pm$ 10.8 &  80.7 $\pm$  5.3 \\
96/05/21  & 224.0897 &  87&.0 $\pm$ 10.6 &  89.8 $\pm$  6.2 \\
96/05/21  & 224.9365 &  94&.5 $\pm$ 11.8 &  88.1 $\pm$  6.6 \\
96/05/23  & 226.0263 &  84&.5 $\pm$ 10.1 &  78.4 $\pm$  5.2 \\
96/05/24  & 227.9291 & 100&.0 $\pm$ 11.9 &  96.1 $\pm$  6.2 \\
96/05/25  & 228.9266 & 102&.8 $\pm$ 12.4 & 100.8 $\pm$  6.8 \\
96/05/26  & 229.9240 &  97&.0 $\pm$ 11.6 &  86.2 $\pm$  5.7 \\
96/06/04  & 238.0026 &  92&.0 $\pm$ 10.8 &  82.2 $\pm$  5.0 \\
96/06/08  & 242.8906 & 104&.0 $\pm$ 12.1 &  99.5 $\pm$  6.0 \\
96/11/23  & 410.4857 & 117&.5 $\pm$ 14.6 & 125.1 $\pm$  9.3 \\
97/05/15  & 583.4548 &  95&.8 $\pm$ 11.3 &  91.5 $\pm$  5.8 \\
\noalign{\smallskip}
\hline
\noalign{\smallskip}
\multicolumn{5}{l}{{\small (a) the automatic analysis results (OLP v7.0) are}} 
\\
\multicolumn{5}{l}{{\small used to compute the photometric error (see text)}} 
\\
\multicolumn{5}{l}{{\small (b) 1$\sigma$ lower limits} } \\
\end{tabular}
\label{tab:lcisocam}
\end{table}

\begin{table}
\caption{ISOPHOT light curves.}
\begin{tabular}{llr@{}rr}
\multicolumn{5}{c}{C1\_60 filter (60 $\mu$m)}  \\ 
\noalign{\smallskip}
\hline
\noalign{\smallskip}
\multicolumn{2}{c}{obs. time} & \multicolumn{2}{c}{flux} &
$\sigma_{var}^{(a)}$ \\
yy/mm/dd  & mjd--50000 & \multicolumn{2}{c}{mJy}  & mJy  \\ 
\noalign{\smallskip}
\hline
\noalign{\smallskip}
96/05/07  & 210.9778 &  430 $\pm$ & 113 &  94 \\
96/05/13  & 216.9627 &  297 $\pm$ &  77 &  64 \\
96/05/15  & 218.9972 &  169 $\pm$ &  66 &  61 \\
96/05/16  & 219.9697 &  581 $\pm$ & 154 & 129 \\
96/05/18  & 221.0536 &  387 $\pm$ & 102 &  86 \\
96/05/18  & 221.9502 &  409 $\pm$ & 215 & 206 \\
96/05/19  & 222.9477 &  404 $\pm$ & 130 & 115 \\
96/05/21  & 224.0960 &  422 $\pm$ & 103 &  83 \\
96/05/21  & 224.9428 &  215 $\pm$ &  80 &  72 \\
96/05/23  & 226.0326 &  519 $\pm$ & 123 &  96 \\
96/05/24  & 227.9355 &  421 $\pm$ &  99 &  78 \\
96/05/26  & 229.9303 &  303 $\pm$ &  91 &  76 \\
96/05/27  & 230.9878 &  286 $\pm$ &  72 &  57 \\
96/06/04  & 238.0089 &  370 $\pm$ &  93 &  75 \\
96/06/08  & 242.8969 &  224 $\pm$ &  71 &  63 \\
96/11/23  & 410.4920 &  315 $\pm$ &  79 &  64 \\
97/05/15  & 583.4611 &  281 $\pm$ &  71 &  59 \\
\noalign{\smallskip}
\hline
\noalign{\smallskip}
\\
\multicolumn{5}{c}{C1\_90 filter (90 $\mu$m)}  \\ 
\noalign{\smallskip}
\hline
\noalign{\smallskip}
\multicolumn{2}{c}{obs. time} & \multicolumn{2}{c}{flux} &
$\sigma_{var}^{(a)}$ \\
yy/mm/dd  & mjd--50000 & \multicolumn{2}{c}{mJy}  & mJy  \\ 
\noalign{\smallskip}
\hline
\noalign{\smallskip}
96/05/07  & 210.9798 &  252 $\pm$ & 102 &  95 \\
96/05/13  & 216.9648 &  342 $\pm$ & 147 & 139 \\
96/05/15  & 218.9992 &  284 $\pm$ & 131 & 124 \\
96/05/16  & 219.9718 &  311 $\pm$ & 159 & 153 \\
96/05/18  & 221.0557 &  192 $\pm$ & 100 &  97 \\
96/05/18  & 221.9522 &  552 $\pm$ & 259 & 245 \\
96/05/19  & 222.9498 &  171 $\pm$ & 102 &  99 \\
96/05/21  & 224.0981 &  273 $\pm$ & 106 &  98 \\
96/05/21  & 224.9448 &  235 $\pm$ & 111 & 105 \\
96/05/23  & 226.0347 &  440 $\pm$ & 200 & 189 \\
96/05/24  & 227.9375 &  299 $\pm$ & 120 & 112 \\
96/05/26  & 229.9323 &  210 $\pm$ & 112 & 106 \\
96/05/27  & 230.9030 &  214 $\pm$ &  83 &  76 \\
96/06/04  & 238.0109 &  262 $\pm$ & 102 &  94 \\
96/06/08  & 242.9024 &  318 $\pm$ & 139 & 131 \\
96/11/23  & 410.4940 &  232 $\pm$ & 129 & 125 \\
97/05/15  & 583.4631 &  267 $\pm$ & 124 & 118 \\
\noalign{\smallskip}
\hline
\noalign{\smallskip}
\multicolumn{5}{l}{{\small (a) error without the responsivity uncertainty}} \\
\end{tabular}
\label{tab:lcisophot}
\end{table}

\begin{figure}[t]
\resizebox{\hsize}{!}{\rotatebox{270}{\includegraphics{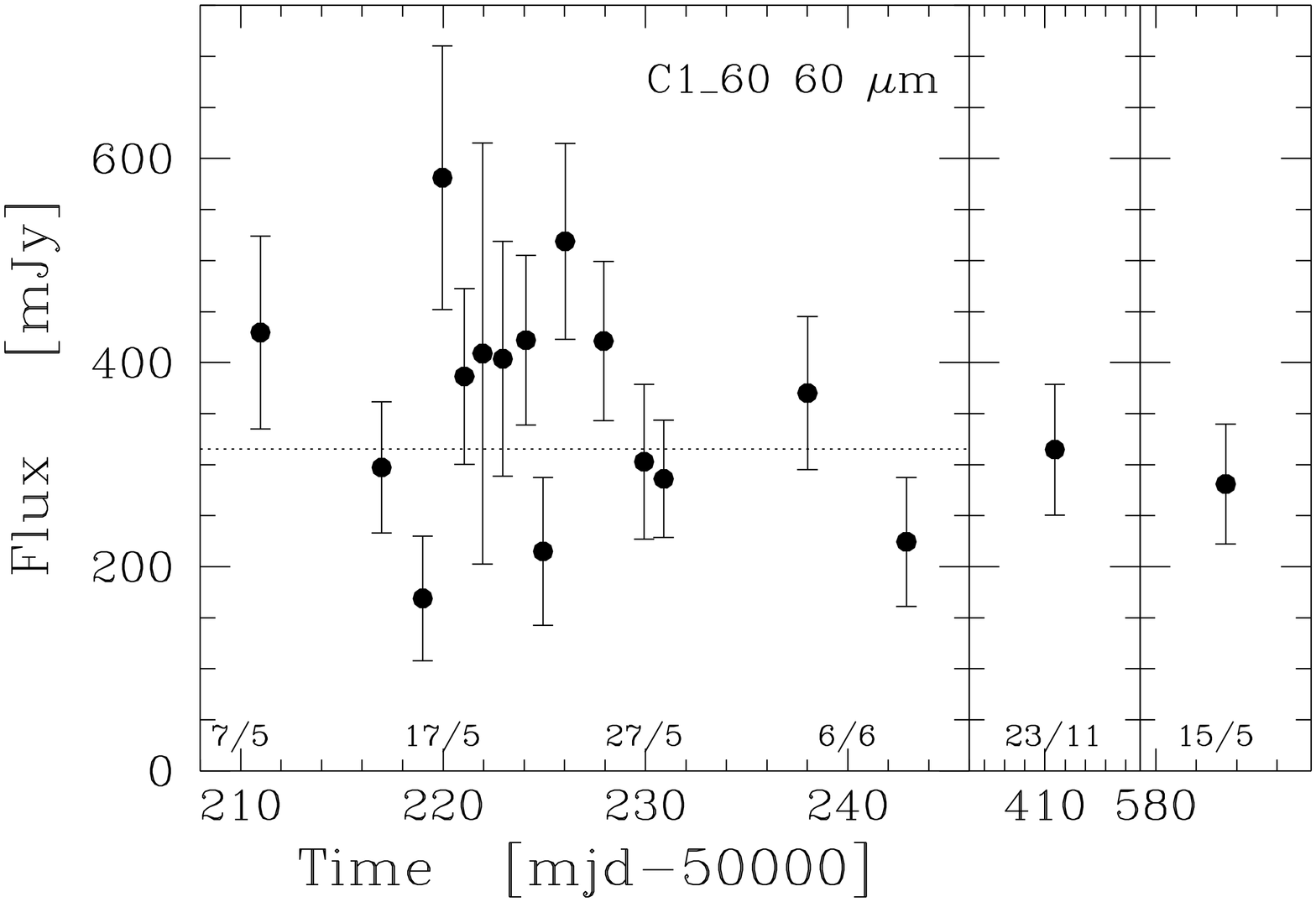}}}
\resizebox{\hsize}{!}{\rotatebox{270}{\includegraphics{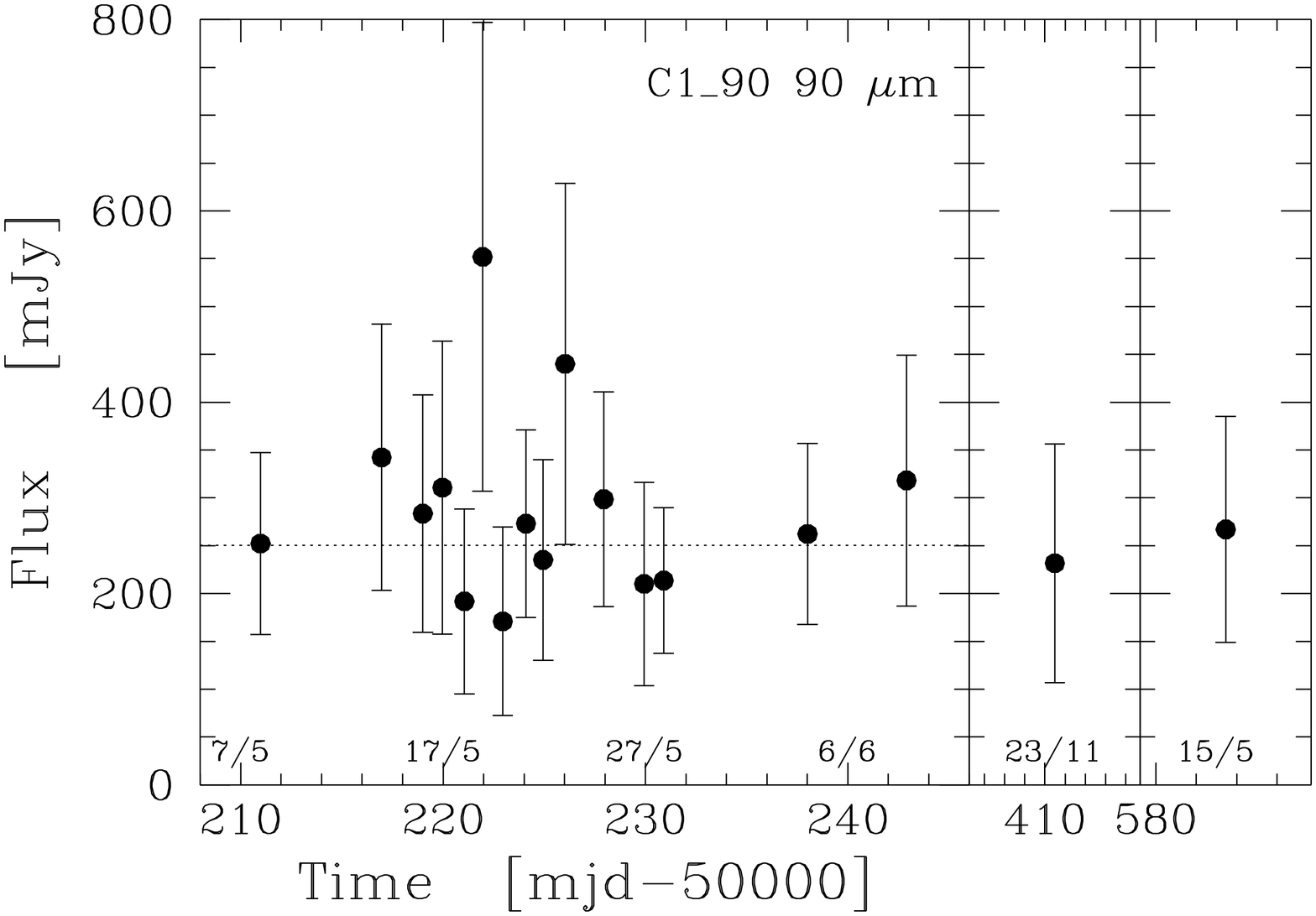}}}
\caption{ISOPHOT light curves of PKS 2155--304. The dotted curves represent the 
fitted constant curves of the best sampled period, from 1996 May 13 to May 27.}
\label{fig:lcisophot}
\end{figure}

When the purpose is to verify whether the flux is variable, the contribution 
of the pixel responsitivity to the absolute error can be neglected and a
smaller uncertainty can be associated to the relative flux values of the light 
curves. 
However, this can be done only for the two light curves of the 
photometer (see Tab. \ref{tab:lcisophot}), due to the way the photometric 
error was determined. 

The relative errors on the flux are, in any case, quite large, about 10 -- 12\%
for the camera observations and from 20 to more than 50\% for the 
photometer (see Appendix A). 
Within these uncertainties the light curves show no 
evidence of variability.
To quantify this statement, we fitted the light curves with a constant term
and the reduced chi--square values were computed in order to test the goodness
of the fits.
We first fitted the values of the best sampled period, from 1996 May 13 to
May 27. The results are $54.8 \pm 1.8$ mJy at 4.0 $\mu$m, $90.8 \pm 3.2$ mJy 
at 14.3 $\mu$m, $315 \pm 27$ mJy at 60 $\mu$m and $250 \pm 34$ mJy at 
90 $\mu$m. To fit the data at 4.0 $\mu$m the lower limits were neglected. We
then repeated the fits, taking the mean of the above--mentioned period and
adding the other data, to look for possible longer--term variability. 
All the fits are acceptable within a confidence level of 95\%. This means
that PKS 2155--304 showed no evidence of variability at these wavelengths 
in the observed period.

However, the large uncertainty on the flux can hide smaller 
variations. We calculated the mean relative error and obtained 3 sigma 
limits for the lowest detectable variations of 32\%, 36\%, 76\% 
and 132\% at 4.0, 14.3, 60 and 90 $\mu$m, respectively.

\subsection{The infrared spectrum}

\begin{figure}[t]
\resizebox{\hsize}{!}{\rotatebox{270}{\includegraphics{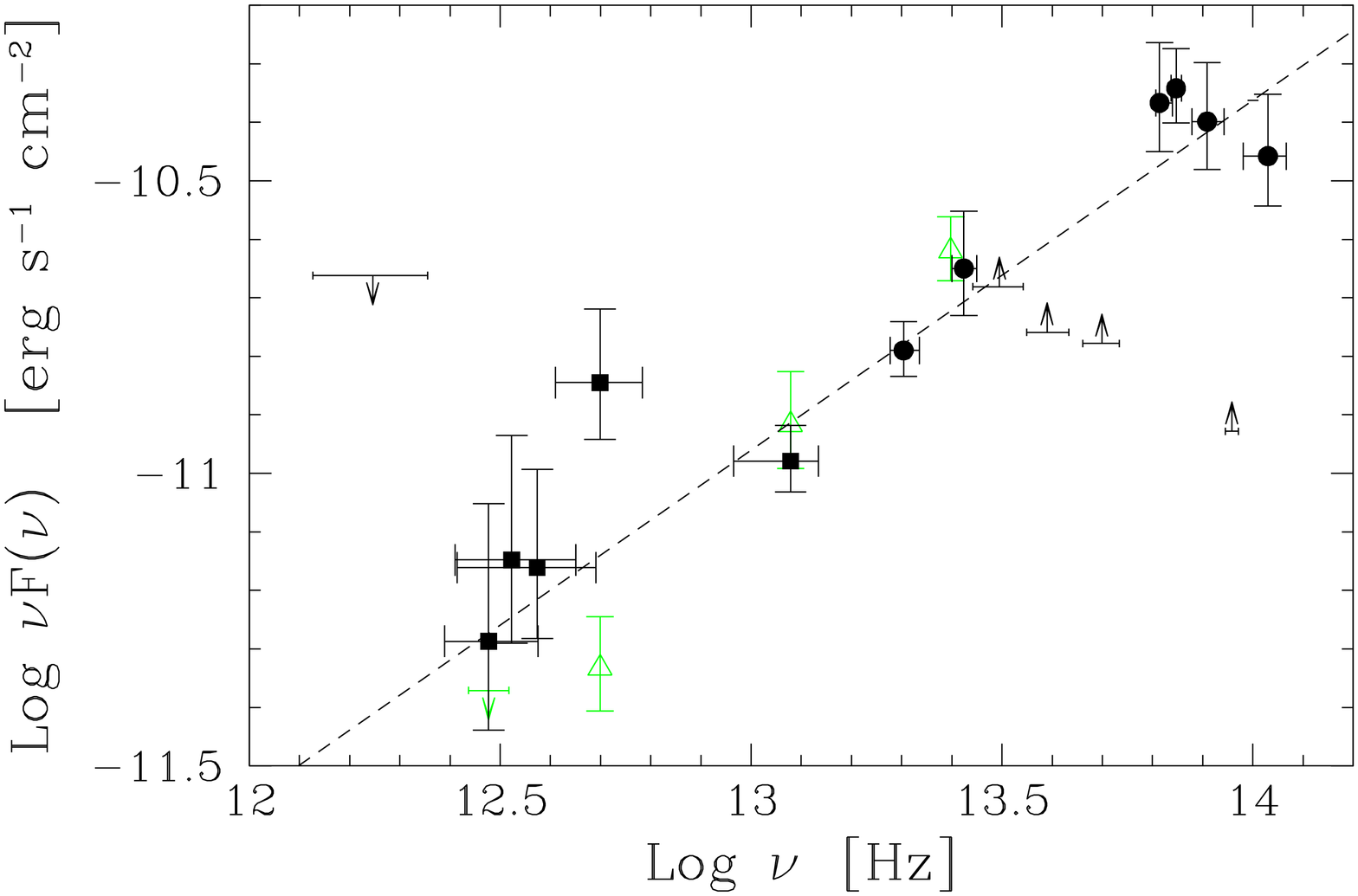}}}
\caption{ISO spectrum. The dashed curve represents the power law fit. Circles 
are the ISOCAM data; squares are the ISOPHOT data. The horizontal bars 
indicate the filter width energy response. The upward arrows are the ISOCAM
lower limits. The downward arrow at
$\log{\nu}\sim12.2$ is the ISOPHOT upper limit at 170 $\mu$m. Open triangles
are the IRAS data (Impey \& Neugebauer \cite{ImpeyNeugebauer})
at 12, 25 and 60 $\mu$m; the IRAS upper limit at 100 $\mu$m is partially
hidden.}
\label{fig:spectrum}
\end{figure}

The infrared spectral shape of PKS 2155--304 was sampled, using 16 filters,
from 2.8 to 170 $\mu$m. The photometer filter C2\_200 was not considered
reliable enough and its observation was discarded. The flux values are given
in Tab. \ref{tab:spectrum} and the spectrum is shown in
Fig. \ref{fig:spectrum}, in a $\log{\nu} - \log{\nu F(\nu)}$ representation.

\begin{table*}
\caption{Fluxes of the observation of 1996 May 27, plus the upper limit 
at 170 $\mu$m.}
\begin{tabular}{lr@{}lcr@{}l@{\hspace{3pt}}l@{}r@{}lr@{\hspace{3pt}}l}
\multicolumn{11}{c}{ISO spectrum}  \\ 
\noalign{\smallskip}
\hline
\noalign{\smallskip}
filter & \multicolumn{2}{c}{$\lambda$} &  $\nu$ &
\multicolumn{5}{c}{flux} & \multicolumn{2}{c}{flux$_{AAR}^{(a)}$} \\ 
 & \multicolumn{2}{c}{$\mu$m} & Hz & \multicolumn{5}{c}{mJy} & 
\multicolumn{2}{c}{mJy} \\ 
\noalign{\smallskip}
\hline
\noalign{\smallskip}
SW4     &  2&.8  & $1.07 \times 10^{14}$ &  32&.5 & $\pm$    &  7&.0 &  38.2 $\pm$ & 3.1 \\
SW2     &  3&.3  & $9.08 \times 10^{13}$ &  13&.0 & $^{(b)}$ &   &   &  28.3 $\pm$ & 5.6 \\
SW6     &  3&.7  & $8.10 \times 10^{13}$ &  49&   & $\pm$    & 10&   &  42.1 $\pm$ & 2.4 \\
SW11    &  4&.26 & $7.04 \times 10^{13}$ &  64&.7 & $\pm$    &  9&.3 &  52.8 $\pm$ & 5.5 \\
SW10    &  4&.6  & $6.52 \times 10^{13}$ &  66&   & $\pm$    & 14&   &  57.1 $\pm$ & 3.9 \\
LW4     &  6&.0  & $5.00 \times 10^{13}$ &  33&.4 & $^{(b)}$ &   &   &  57.1 $\pm$ & 2.8 \\
LW6     &  7&.7  & $3.89 \times 10^{13}$ &  44&.7 & $^{(b)}$ &   &   &  61.5 $\pm$ & 2.6 \\
LW7     &  9&.6  & $3.12 \times 10^{13}$ &  66&.7 & $^{(b)}$ &   &   &  80.6 $\pm$ & 2.4 \\
LW8     & 11&.3  & $2.65 \times 10^{13}$ &  84&   & $\pm$    & 17&   & 122.0 $\pm$ & 3.7 \\
LW9     & 14&.9  & $2.01 \times 10^{13}$ &  80&.6 & $\pm$    &  8&.6 &  83.0 $\pm$ & 3.1 \\
P2\_25  & 25&    & $1.20 \times 10^{13}$ &  88&   & $\pm$    & 11&   &       & \\
C1\_60  & 60&    & $5.00 \times 10^{12}$ & 286&   & $\pm$    & 72&   &       & \\
C1\_70  & 80&    & $3.75 \times 10^{12}$ & 184&   & $\pm$    & 59&   &       & \\
C1\_90  & 90&    & $3.33 \times 10^{12}$ & 213&   & $\pm$    & 83&   &       & \\
C1\_100 &100&    & $3.00 \times 10^{12}$ & 172&   & $\pm$    & 72&   &       & \\
C2\_160 &170&    & $1.76 \times 10^{12}$ &1235&   & $^{(c)}$ &   &   &       & \\
\noalign{\smallskip}
\hline
\noalign{\smallskip}
\multicolumn{11}{l}{{\small (a) the automatic analysis results (OLP v6.3.2)
are used to}} \\
\multicolumn{11}{l}{{\small compute the photometric error (see text)}}
\\
\multicolumn{11}{l}{{\small (b) 1$\sigma$ lower limit}} \\
\multicolumn{11}{l}{{\small (c) 3$\sigma$ upper limit}}
\end{tabular}
\label{tab:spectrum}
\end{table*}

In Fig. \ref{fig:spectrum} it is also shown the result of a power law fit, 
that gives an energy spectral index of 
$\alpha = 0.40 \pm 0.06$. The lower and upper limits
were not considered in the fit; the reduced chi--square is
$\chi^2_r = 1.31$, with 9 d.o.f., that gives a confidence level of 
77.4\%.

 From each simultaneous pairs of flux values of the SW5 and LW3 light curves, 
we obtained the spectral indices between 4.0 and 14.3 $\mu$m as 
$\alpha_i = - \log(f_{SW5,i}/f_{LW3,i})/\log(\nu_{SW5}/\nu_{LW3})$.
The mean value is $\alpha = 0.403 \pm 0.017$, which is fully consistent 
with the index derived using 11 filters on a larger IR band.

The fit with a constant term of the spectral indices $\alpha_i$ vs. time has 
a reduced chi--square of 0.26, with 13 d.o.f., which corresponds to a 
confidence level of less than 1\%. This indicates that the source showed no 
spectral variability in the 4.0 -- 14.3 $\mu$m range, during the observed 
period.


\section{Optical observations}
\label{sec:optobs}

\subsection{Observations and data reduction}

The optical data were obtained using the Dutch 0.9 m ESO 
telescope at La Silla, Chile, between May 17 and 27 1996. 
The telescope was equipped with a TEK CCD 512x512 pixels 
detector and Bessel BVR filters were used for the observations. 
The pixel size is 27 $\mu$m and the projected pixel 
size in the plane of the sky is 0.442 arcsec, providing a field of view 
of 3$^{\prime}$.77 x 3$^{\prime}$.77.

The original frames were flat fielded and bias corrected using
MIDAS package and photometry was performed using the Robin procedure, 
developed at the Torino Observatory, Italy, by L. Lanteri. 
This procedure fits the PSF with a circular 
gaussian and evaluates the background level by fitting it with 
a 1st order polynomial surface. The magnitude of the object and the error 
are derived by comparison with reference stars in the same field 
of view. The typical photometric error is $\sim 0.02$ mag in all
bands.

\subsection{Results}

\begin{figure}
\resizebox{\hsize}{!}{\includegraphics{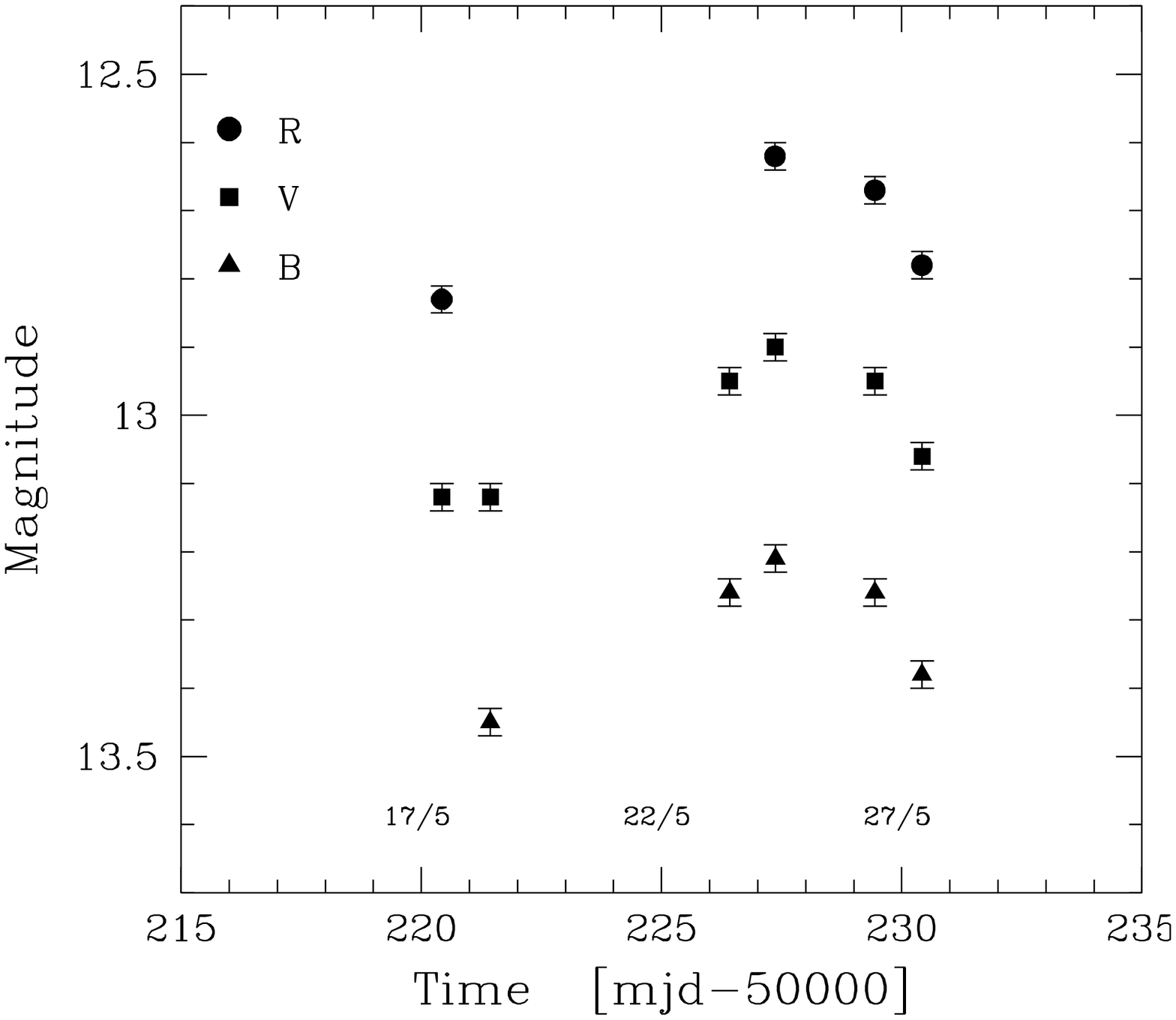}}
\caption{Optical light curves of PKS 2155--304.}
\label{fig:lcoptical}
\end{figure}

The light curves (Tab. \ref{tab:lcoptical} and Fig. \ref{fig:lcoptical}) 
show an increase of luminosity of about 20\% 
($\sim 0.2$--$0.25$ mag), between the starting low level of May 17--18 and 
the maximum of May 24. The flux is then decreasing during the last two days.
The behavior is very similar in all of the three filters.

Assuming that the optical spectrum is described by a power law, we 
calculated the mean spectral indices using the simultaneous pairs data 
of the light curves. The results are $\alpha_{RV} = 0.62 \pm 0.02$ and
$\alpha_{VB} = 0.60 \pm 0.02$ and indicate that the optical spectrum is 
steeper than the IR one.

\begin{table}
\caption{RVB light curves.}
\begin{tabular}{llccc}
\noalign{\smallskip}
\hline
\noalign{\smallskip}
\multicolumn{2}{c} {obs. time} &  R & V & B \\
yy/mm/dd  & mjd--50000 & mag & mag & mag \\ 
\noalign{\smallskip}
\hline
\noalign{\smallskip}
96/05/17  & 220.4250 & 12.83 &  ...  &  ...  \\
          & 220.4277 &  ...  & 13.12 &  ...  \\
96/05/18  & 221.4298 &  ...  &  ...  & 13.45 \\
          & 221.4354 &  ...  & 13.12 &  ...  \\
96/05/23  & 226.4146 &  ...  & 12.95 &  ...  \\
          & 226.4160 &  ...  &  ...  & 13.26 \\
96/05/24  & 227.3625 & 12.62 &  ...  &  ...  \\
          & 227.3660 &  ...  & 12.90 &  ...  \\
          & 227.3688 &  ...  &  ...  & 13.21 \\
96/05/26  & 229.4396 & 12.67 &  ...  &  ...  \\
          & 229.4409 &  ...  & 12.95 &  ...  \\
          & 229.4417 &  ...  &  ...  & 13.26 \\
96/05/27  & 230.4194 & 12.78 &  ...  &  ...  \\
          & 230.4202 &  ...  & 13.06 &  ...  \\
          & 230.4215 &  ...  &  ...  & 13.38 \\
\noalign{\smallskip}
\hline
\noalign{\smallskip}
\multicolumn{5}{l}{{\small Note: for all bands uncertainties are $\sim 0.02$ mag}} \\
\end{tabular}
\label{tab:lcoptical}
\end{table}

\section{Discussion}
\label{sec:discussion}

\subsection{IR flux and spectral variability}

The ISO light curves of May--June 1996 show that the time variability of PKS
2155--304 in the mid-- and far--infrared bands is very low or even absent. 
The flux has not varied significantly in 1996 November and in 1997 May, 
one year later, and is quite similar to the 1983 IRAS state 
(Impey \& Neugebauer \cite{ImpeyNeugebauer}) (Fig. \ref{fig:spectrum}), except at 60 $\mu$m, 
where the IRAS flux seems significantly lower.
This agreement could support the idea that the infrared flux level of
this source is rather stable. 
We have to wait for future satellite missions to test this statement.

The infrared spectrum from 2.8 to 100 $\mu$m is well fitted by a single power
law. This is a typical signature of synchrotron radiation, that can
explain the whole emission in this wavelength range, excluding 
important contributions of thermal sources.

The variability in the optical bands is small too, while the simultaneous RXTE
light curve (Urry et al. \cite{Urry98}, \cite{Sambruna}) shows,
on the contrary, strong and fast variability at energies of 2--20 keV: the
flux varied by a factor 2 on a timescale shorter than a day. 
This seems to be a common behavior in blazars, for which there is a 
more pronounced variability at frequencies above the synchrotron peak 
(\cite{UlrichMaraschiUrry}).

\subsection{Contribution of the host galaxy to the IR flux}

The absence of variability could be also explained by the contribution, in the
IR, of a steady component, such as the host galaxy.
The host galaxy of PKS 2155--304 is a large elliptical which is well 
resolved in near infrared images (\cite{KotilainenFalomoScarpa}), 
but the pixel field of view of the ISOCAM camera (3\arcsec or 6\arcsec) is 
too big to resolve it and its contribution is integrated in the flux 
of the active nucleus.

The magnitude of the host galaxy in the $H$ band is $m_H = 12.4$
(\cite{KotilainenFalomoScarpa}).
The color of a typical elliptical at $z=0.11$ is $B-H$=4.6 (\cite{Buzzoni}), 
from which we get $m_B = 17.0$, which corresponds to a flux $f_B$ = 0.7 mJy. 
Mazzei \& De Zotti (\cite{MazzeiDeZotti}) calculated the flux ratio between the IRAS and the 
$B$ bands for a sample of 47 elliptical galaxies: their results are 
$\log{f_{12}/f_B} = 0.01 \pm 0.05$, 
$\log{f_{25}/f_B} = -0.70 \pm 0.32$, 
$\log{f_{60}/f_B} = -0.22 \pm 0.155$, 
$\log{f_{100}/f_B} = 0.25 \pm 0.10$. 
 From these relations we can estimate the host galaxy fluxes in the 
far--IR at 12, 25, 60 and 100 $\mu$m: 
we have 
$f_{12} = 0.7$ mJy, 
$f_{25} = 0.1$ mJy,
$f_{60} = 0.4$ mJy, 
$f_{100} = 1.2$ mJy. 
If we compare these values with those of Tab. \ref{tab:spectrum}, we see
that they are less than 1\% of the active nucleus flux, and much less than
the uncertainties.
We thus conclude that the contribution of the host galaxy to the ISO far--IR 
flux is negligible.

\begin{figure}
\resizebox{\hsize}{!}{\includegraphics{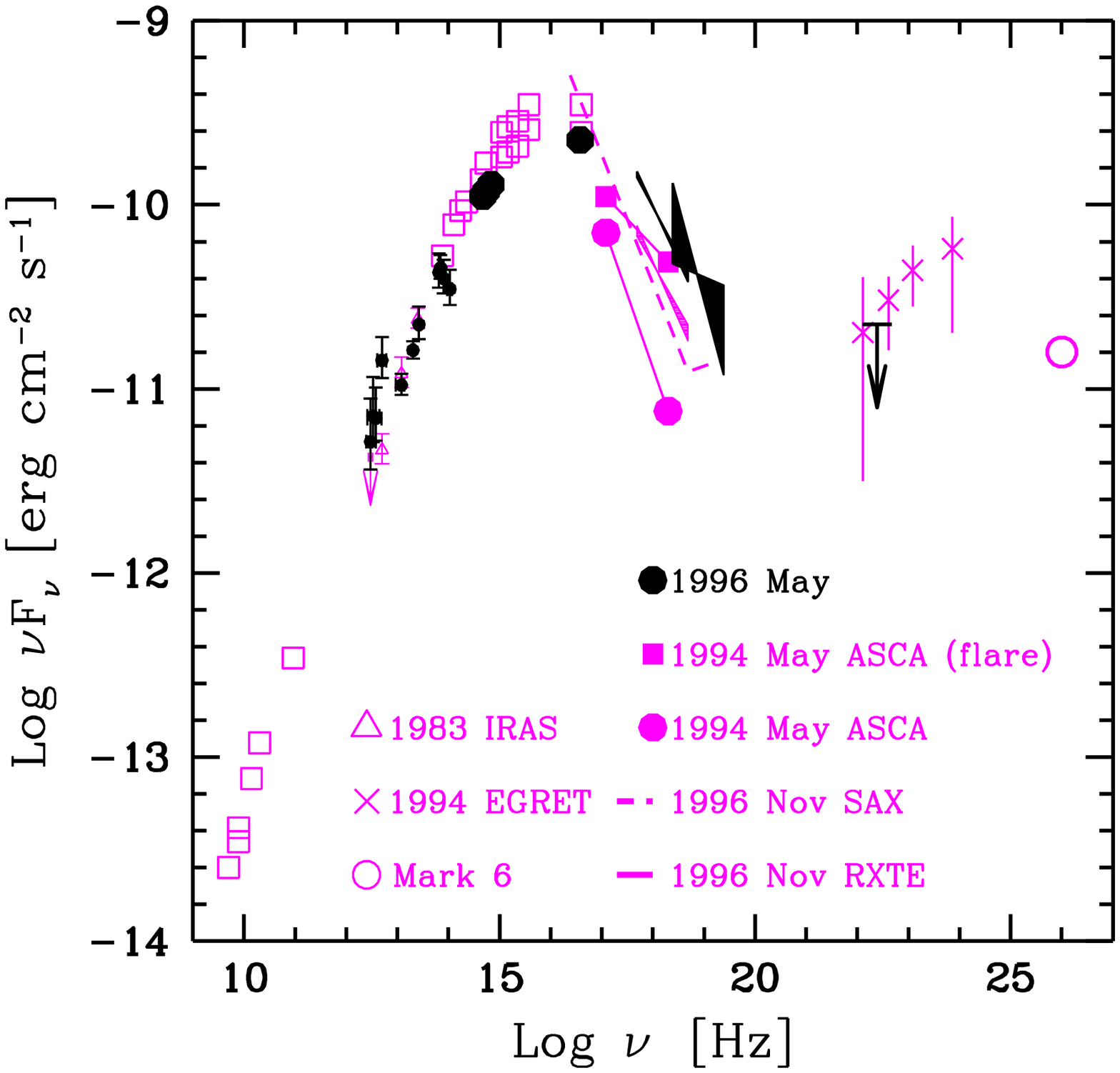}}
\caption{Spectral energy distribution of PKS 2155--304. The black data are
the simultaneous data of 1996 May: circles are ISO and BVR fluxes (this paper, 
data of May 27), and EUVE flux (Marshall H.L., priv. comm.); the X--ray 
spectra  are from RXTE (Urry et al. \cite{Urry98}); the $\gamma$--ray upper limit is from 
EGRET (Vestrand W.T., priv. comm.). Open grey boxes are data from the
multiwavelength campaign of 1994 May (Pesce et al. \cite{Pesce}, Pian et al. \cite{Pian},
Urry et al. \cite{Urry97}). 1994 ASCA data are from Urry et al. (\cite{Urry97}), 1994 EGRET data are
from Vestrand et al. (\cite{Vestrand}), 1996 SAX spectrum is from Giommi et al. (\cite{Giommi}), 1996 November
RXTE spectrum is from Urry et al. (\cite{Urry98}), IRAS data are from Impey \& 
Neugebauer (\cite{ImpeyNeugebauer}), Mark 6 point is from Chadwick et al. (\cite{Chadwick}).}
\label{fig:SED}
\end{figure}

This fact can be also inferred from the spectral energy distribution (SED), 
built with the simultaneous data of May 1996 (Fig. \ref{fig:SED}), that shows
that the ISO data lie on the interpolation between radio and optical
spectra.

\subsection{Synchrotron self--absorption}

The observed IR spectrum is rather flat, and one can wonder if this is due
to a partially opaque emission, i.e. if we have, in the IR, the superposition
of components with different self--absorption frequencies,
as for the flat radio spectra. 

To show that this is $not$ the case, we calculate the self--absorption
frequency assuming that the IR radiation originates in the same compact
region responsible for most of the emission, including the strongly
variable X--ray flux.
This is a conservative assumption, since the more compact is the region,
the larger is the self--absorption frequency.
In the case of an isotropic population of relativistic electrons with a
power--law distribution $N(\gamma) = K \gamma^{-p}$, 
the self--absorption frequency is given by (e.g. Krolik 1999)
\begin{displaymath}
\nu_t = {\delta \nu_B \over 1+z} \, 
\left[ {3^{p \over 2} \pi \sqrt{3 \pi} \over 4} \,
{\Gamma ({3p+22\over 12}) \Gamma ({3p+2\over 12}) 
\Gamma ({p+6\over 4}) \over \Gamma ({p+8 \over4}) }\, 
{e \tau \over B \sigma_T}
\right]^{\frac{2}{p+4}},
\nonumber
\end{displaymath}
\noindent
where $\Gamma$ is the gamma function, $\nu_B$ is the cyclotron frequency, 
$\delta$ is the beaming factor, $R$ is the size of the source,
$\tau \equiv \sigma_T K R$, and $p$ is the slope of the electron distribution
appropriate for those electrons radiating at the self--absorption energy.
In the homogeneous synchrotron self--Compton model, the optical depth 
$\tau$ is approximately the ratio of the Compton and synchrotron flux at 
the same frequency. 
This ratio can be estimated from the SED (Fig. \ref{fig:SED}), where
the Compton flux is obtained by extending at low frequencies the 
Compton spectrum with the same spectral index of the synchrotron curve. 
The upper limit for the $\gamma$--ray emission in 1996 May corresponds  
to an upper limit for the value of the optical depth of $\tau \lesssim
10^{-5}$. From the ISO spectrum, we have $p = 2\alpha + 1 = 1.8$. 
Although we cannot a priori determine the other two parameters, 
namely $B$ and $\delta$, a reasonable estimate can be derived through 
the broad band model fitting. In particular if we adopt the 
values derived by Tavecchio et al. (\cite{Tavecchio}), 
$B = 0.25$ G and  $\delta \sim 30$, we get 
$\nu_t \lesssim 1.4 \times 10^{11}$ Hz. 
For less extreme values of $\delta$, $\nu_t$ becomes smaller,
while much larger values of the magnetic field 
(making $\nu_t$ to increase) are implausible,
if the significant $\gamma$--ray emission is due to the self--Compton process,
which requires the source not to be strongly magnetically dominated.
The frequency of self--absorption is thus significantly lower then the IR 
frequencies, implying that the IR emission is completely thin.

\subsection{Spectral energy distribution}

In Fig. \ref{fig:SED} we show the SED of PKS 2155--304 during our
multiwavelength campaign, from the far IR to the $\gamma$--ray band.
We also collected other, not simultaneous, data from the literature, especially
in the X--ray band, to compare our overall spectrum with previous observations.
As can be seen, our IR data fill a hole in the SED and, together with our
optical results, contribute to a precise definition of the shape of the
synchrotron peak.
It is remarkable that although the X--ray state during our campaign was very 
high (one of the highest ever seen), the optical emission was not
particularly bright.
Also the upper limit in the $\gamma$--ray band testifies that the source
was not bright in this band.

All this can be explained assuming that the X--ray flux is due to the steep
tail of an electron population distributed in energy as a broken power law.
The first part of this distribution is flat and steadier than the
high energy, steeper part.
In this case {\it without changing significantly the bolometric 
luminosity} large flux
variations are possible above the synchrotron (and the Compton) peak.
An electron distribution with these characteristics can be obtained
by continuous injection and rapid cooling (see e.g. \cite{Ghisellini}).
In fact, if the electrons are injected at a rate 
$Q(\gamma)\propto \gamma^{-s}$
between $\gamma_1$ and $\gamma_2$, the steady particle distribution
will be $N(\gamma)\propto \gamma^{-(s+1)}$ above $\gamma_1$, and
$\propto \gamma^{-2}$ below, until radiation losses dominate the 
particle escape or other
cooling terms (e.g. adiabatic expansion).
Electrons with energy $\gamma_1 m_ec^2$ are the ones responsible for the 
emission at the synchrotron and Compton peaks (as long as the scattering 
process is in the Thomson limit).
Since it is possible to change $s$ without changing the total injected power,
large flux variations above the peak are compatible with only
minor changes below.
This model also predicts that the spectrum below the peak has a slope 
$\alpha=0.5$, which is not far from what we have observed in the far IR.


\begin{acknowledgements}
We would like to thank the ISOCAM team and, in particular, Marc Sauvage for 
his help with CIA, the ISOCAM data reduction procedure and with the 
installation of the software at OAB.
We also thank Giuseppe Massone e Roberto Casalegno, who made the optical 
observations at La Silla.
\end{acknowledgements}


\appendix

\section{Data reduction}
\label{sec:datareduction}

\subsection{ISOCAM}

The observations were processed with CIA\footnote{ISOCAM Interactive Analysis,
CIA, is a joint development by the ESA Astrophysics Division and the
ISOCAM Consortium led by the ISOCAM PI, C. C\'esarsky, Direction des
Sciences de la Mati\`ere, C.E.A., France.} v2.0.

Each observation consisted in a sequence of frames, which had an
elementary integration time of about 2 s. By this way the temporal behaviour
of each pixel was known.

First, the dark current was subtracted from each raw frame, using the dark
images present in the software library, flagging the bad pixels of SW and LW
detectors.

The impact of charged particles (glitches) on the detectors create spikes in
the pixel signal curve. To remove these spurious signals, we first used the 
Multiresolution Median Transform method (\cite{Starck}),
then every frame was inspected to make sure that the number of suppressed
noise signals was negligible and finally a manual deglitching operation was
done to detect the glitches left and flag them.
Some glitches caused a change in the pixel sensitivity: in this case we
flagged the pixel in all readouts after the glitch.

The library dark images were not good enough to remove all the effects
of the dark current: the signals in rows and columns showed a saw--teeth
structure, that was eliminated using the Fast Fourier Transform technique
(\cite{StarckPantin}).

The response of the detector pixels to a change in the incident flux is
not immediate and the signal reaches the stabilization after some time.
This time interval depends on the initial and final flux values and on 
the number of readouts (\cite{ISOCAMobsman}). Therefore, the time sequence 
of a pixel signal shows, after a change in the incident flux, an upward 
or downward transient behaviour. 
At the beginning of every observation, after a certain number of 
frames, the signal should reach the stable value. As this 
ideal situation could not always been achieved, CIA provides different 
routines to overcome this problem and apply the transient correction. These
routines use different models to fit the signal curves, in order to identify
the stable value.

In the SW5 observations, the photons coming from PKS 2155--304 fell mainly 
in one or two pixels, whose signals showed an upward transient behaviour that
never reached the stabilization. On the contrary, the background, being very
low, was stabilized. No transient correction routines were able to adequately
fit the source signals, either underestimating or overestimating the stable 
flux.
Observing the signal curves, we noticed that the behaviour of the first part 
of the curves were far from the expected converging trends that are used in 
the models of the correction routines, while the remaining part of the curves
seemed to be well described by a converging exponential trend.
So, after having discarded the starting readouts, we fitted the signal with a
simple exponential model $s_{fit} =  s_{\infty} +  c \cdot e^{-t/\tau}$,
where the optimized parameters are $c$, $\tau$ and $s_{\infty}$, that 
represents the stable signal.
We chose the fit which showed a reasonable result and optimized
the determination coefficient
$R^2 = 1 - \frac {\sum_i (s_i - s_{fit})^2}{\sum_i (s_i - \bar{s})^2}$,
where $s_i$ are the measured signals and $\bar{s}$ is the mean of the
part we considered.
In three cases, the results were not acceptable and we could define only
lower limits, as the upward transients had not reached the stabilization.

For the transient correction of the LW3 observations, the model developed at 
the {\it Institut d'Astrophysique Spatiale} (IAS Model)
(\cite{AbergelDesertAussel}) has been used. As the corrected curves attained
stable values in the second half only, we did not use the first half of the
frames.

In the spectral observation of May 27, the uninterrupted sequence of filters 
used created either upward or downward transients and the stabilization of the 
source signal was reached just in few cases. 
The five observations made with the SW channel were corrected using the same 
method as the SW5 ones, except for the SW11 filter, in which the stabilization 
was reached for all pixels. In this case, we just discarded the first half of
the 162 frames. In the SW2 filter data, at the end of the observation, the
source signal was so far from stabilization that we could define only a lower
limit.
The five observations made with the LW channel were corrected using the IAS 
model. As this model takes into account all the past illumination history, we 
fitted a unique curve that was built linking together all the LW filters data.
This method worked fine for two filters only (LW8 and LW9), while for the 
other three filters again we defined lower limits.

We averaged all the frames neglecting the flagged signal values and then the 
images were flat fielded, using the library flat fields of CIA.

The total signal of the source was computed integrating the values of
the signal in a box centered on the source and subtracting the normalized
background obtained in a ring of 1 pixel width around the box. The boxes 
had dimension ranging from 3x3 to 7x7 pixels, depending on the filter and on 
the pixel field of view (pfov).
The results were colour corrected and divided for the point spread function 
(PSF) fraction falling in the box. This fraction also depends on filters and 
pfov. To compute it, we extracted from the library, for each 
combination of filter and pfov, the nine PSF images centered more or less on
the same pixels of PKS 2155--304.
For calibration requirements, in each PSF image the centroid of the source 
was placed in a slightly different position inside the same pixel.
As we do not know with enough accuracy the position of the centroid of 
PKS 2155--304 in the ISOCAM images, the nine PSF were averaged and the result
was normalized. 
The PSF correction was calculated by summing the signal of the pixels in 
a box of the same dimension of that in which we extracted the source signal.
For the LW detector, a further correction factor was applied to take into 
account the flux of the point--like source that falls outside the
detector (\cite{Okumura}). For the SW channel, we adopted for all filters
the SW5 PSF, because, along with SW1, was the only one present in the 
calibration library, however the error we introduced can only be of a few 
percent.

Finally, the source signal was converted to flux density using the coefficients
in Blommaert (\cite{Blommaert}).

To compute the photometric error we divided the uncertainty sources in 
two parts: the first one took into account the dark current
subtraction, deglitching,
flat fielding operations and signal to flux conversion, while the second one 
considered the transient correction.
The first group of error sources are derived from 
the Automatic Analysis Results (AAR; OLP v7.0 for the light 
curves data, OLP v6.3.2 for the spectrum data). The source flux values
$f_{AAR}$ given by the AAR are not reliable because the transient correction
is not performed, but the AAR absolute flux errors $\sigma_{AAR}$ are a good
estimate of the first group of errors (the
AAR fluxes are given in Tabs. \ref{tab:lcisocam} and \ref{tab:spectrum}). We
assumed that the fluxes $f_{src}$ that we derived have the same relative 
error $\sigma_{rel} = \sigma_{AAR}/f_{AAR}$. Thus, for our fluxes this part of 
error is $\sigma_f = \sigma_{rel} \, f$, which accounts for all the 
uncertainties sources, but the transient correction. We estimated that the 
error due to the transient correction is of the order of 10\%, 
which is the rounded maximum error on the stable signal
$s_{\infty}$, obtaining a total error of $\sigma = \sqrt{\sigma^2_f +
\sigma^2_{tr}}$. We assumed then a $\sigma_{tr}$ of 10\% 
for all our measurements (20\% for SW4 and SW10 filters).

\subsection{ISOPHOT}
\label{sec:PHOTdatareduction}
         
The observations were done in rectangular chopped mode: the observed field
of view switches alternately between the source and an 180\arcsec\,distant
off--source position. This is necessary in order to measure the background
level. The chopping direction was along the satellite Z-axis, which was slowly 
rotating by about one degree per day. Thus, every time the background was
sampled in different fields of the sky and a raster map was performed just 
to check the stability of the background all around the source. The standard 
deviation of the background flux measured in the
central pixel of the C100 detector, in the eight off-source positions of the
scan, is 37 mJy. This value is much less than the error of the source flux
(see Tab. \ref{tab:lcisophot}). This small background fluctuation 
would lead to a rise of the scatter of the source flux, in any case our 
results are compatible with absence of variability (see section
\ref{sec:ISOresults}).

Each observation of an astronomical target was immediately followed by
a Fine Calibration Source (FCS) measurement, using internal calibrations 
sources. These measurements were made in order to determine the 
detector responsitivity, which is necessary to compute the target flux.

Each observation consisted in a series of integration ramps, each one made 
by the sequence of voltage readouts between two destructive readouts.

The observations were processed with PIA\footnote{ISOPHOT Interactive
Analysis (PIA) is a joint development by the ESA Astrophysics Division and
the ISOPHOT Consortium. The ISOPHOT Consortium is led by the
Max--Planck--Institute for Astronomy, Heidelberg.} v7.0 (\cite{Gabriel}).

PIA separates the operations to be performed on the data in different levels:
at each level PIA creates a data structure on which it operates. This data
structure takes its name according to the properties of the data.
The first part of the data analysis was common for all the observations, then
the procedures changed according to the different characteristics of the
observation (whether it was chopped or not or whether the detector was
receiving photons from the astronomical target or from the FCS).

At the beginning, PIA automatically converted the digital data from telemetry
in meaningful physical units and created the structure of data, called Edited
Raw Data (ERD).
At the ERD level, some starting readouts and the last readout
of each ramp were discarded, because they are disturbed by the voltage 
resetting; we also manually discarded the part of the ramp before or after 
a glitch (that causes a sudden jump of the readout value) in the cases 
where most part of the ramp was unchanged and the glitch did not modify the 
detector responsitivity. A correction for the non--linear responsitivity of the
detector was applied, using special calibration files. Then, each ramp was
fitted by a 1st order polynomial model. A signal (in V s$^{-1}$) was obtained
from the slope of every ramp: the slope is proportional to the incident power.
At Signal per Ramp Data (SRD) level, the first half of the signals per chopper 
plateau were discarded, because of stabilization problems. As the signal value
depends on the integration time, a correction factor was applied and the 
signal was normalized for an integration time of 1/4 s. 
The dark current was subtracted using the PIA calibration files, which take 
into account the satellite position in the orbit.
An algorithm was applied to discover and 
discard the signals that were anomalously high, because of glitches; then, 
the signals of each chopper plateau were averaged.
At Signal per Chopper Plateau (SCP) level, the responsitivity of each 
detector pixel was computed taking the median of the FCS2 signals of the 
calibration measurements; then, the vignetting correction was performed on 
the target observations. In the chopped measurements, the background, that 
was calculated at the off--source position, was subtracted to get the source 
signal.

As for the camera, the response of the photometer detectors has some delays 
after a change in the incident flux. This effect causes losses in the 
signal values measured in the chopped measurements, so a correction factor 
was applied. The signal was finally converted into power, using the 
responsitivity obtained from the FCS measurement.

In the observations performed with the 3x3 pixel C100 detector, only the 
central pixel was used to compute the source flux density, because, as the
most of the Airy disk of a point-like source centered in the pixel lies in the 
same pixel (69\% for C1\_60 and 61\% for C1\_90), to use the outer pixels 
just adds more noise than signal.
The source flux density is defined as
$F_{\lambda} = P_{src}/(C1 f_{psf})$, where $P_{src}$ is the incident power,
$C1$ is a conversion factor of each filter (as given in the PIA calibration
file pfluxconv.fits) and $f_{psf}$ is the fraction of PSF that falls on the
pixel considered when the source is located in the centre
(\cite{ISOPHOTobsman}, Tabs. 2 and 4).

The absolute photometric error was computed by PIA, during the data reduction
process, and took into account the uncertainty in the determination of the
slope of the ramp and the errors associated to the other performed correction
operations.



\end{document}